\documentclass[reprint,amsmath,amssymb,aps,]{revtex4-1}
\usepackage{dcolumn}
\usepackage{bm}
\usepackage{amsmath}
\usepackage{graphicx}
\usepackage{epstopdf}
\usepackage[caption=false]{subfig}
\begin{document}

\title{Efficiency of a two-stage heat engine at optimal power}

\author{I. Iyyappan}
\email[]{iyyappan@iisermohali.ac.in}
\author{Ramandeep S. Johal}
\email[]{rsjohal@iisermohali.ac.in}
\affiliation{Department of Physical Sciences,\\ Indian Institute of Science Education and Research Mohali,\\
Sector 81, S. A. S. Nagar, Manauli PO 140306, Punjab, India.}
\begin{abstract}
We propose a two-stage cycle for an optimized linear-irreversible heat engine that operates, in a finite time,  between a hot (cold) reservoir and a finite auxiliary system acting as a sink (source) in the first (second) stage. Under the tight-coupling condition, the engine shows the low-dissipation behavior in each stage, i.e. the entropy generated depends inversely on the duration of the process. The phenomenological dissipation constants 
are determined within the theory itself in terms of the heat transfer coefficients and the heat capacity of the auxiliary system. We study the efficiency at maximum power and highlight a class of efficiencies in the symmetric case that show universality upto second order in Carnot efficiency, while Curzon-Ahlborn efficiency is obtained as the lower bound for this class.
\end{abstract}
\maketitle
 
The central result in  thermodynamics is that a reversible heat cycle between a hot and a cold reservoir achieves the Carnot efficiency \cite{carn}: 
\begin{equation}
\eta_{\rm C} = 1-\frac{T_c}{T_h},
\end{equation}
where the temperatures of the reservoirs satisfy: $T_{c} <T_h$. This result is outstanding for its independence from the nature of the working medium and the exact type of reversible cycle. Real-world machines mostly operate in irreversible regimes with efficiencies that are system and/or mechanism dependent \cite{cur22,and157,and208,rub127,sal211,and62,dev570,che374,ang746,che114,bej119,che99,sal354}. Thus, it is remarkable to find universal features for efficiency in different 
models of irreversible machines \cite{van190,sch200,tu312,esp130,esp150,ben230,izu100,bra070,hol050,tu052,vau199,jia042,bau042,bran031,pro220,proe041,rya050,joh0121,dec500,joh500}. 

Various models have been proposed in the literature, such as the endoreversible  \cite{cur22}, stochastic \cite{sch200,dec500} and the low-dissipation (LD) \cite{esp150}, which work with a finite cycle time. Steady-state thermal machines have been studied using linear-irreversible framework \cite{van190} and beyond \cite{izu100}. Although the formula rediscovered by Curzon and Ahlborn \cite{cur22,vau199}, $\eta_{\rm CA}^{} = 1-\sqrt{1-\eta_{\rm C}^{}}$, was initially suspected to be a generic result for  
efficiency at maximum power (EMP), it actually applies to small temperature gradients (small values of $\eta_{\rm C}$), whereby $\eta_{\rm CA}^{} \approx {\eta_{\rm C}}/{2} + {\eta_{\rm C}^{2}}/{8}+ O[{\eta_{\rm C}^{3}}]$. The universal nature of the first and the second-order terms have been subsequently elaborated in various models \cite{sch200,tu312,esp130}.
Apart from finite-rate mechanisms, models with finite-size reservoir(s) have
also been studied \cite{izu180,wan062,joh100,joh012}.

In this Letter, we propose a two-stage linear-irreversible heat engine which cyclically runs between an infinite-size hot (cold) reservoir and a finite-size auxiliary {\it system} serving as a heat sink (source) in the first (second) stage. Our first main result shows that
a linear-irreversible engine with a finite-size sink or source, and optimized for its performance in a given time, shows the  LD behavior such that the total entropy generated in the process is inversely proportional to the duration of the process \cite{esp150}. More precisely, LD behaviour is obtained as a reasonable approximation to our model in the strong-coupling limit. The second main result is that, under a condition of symmetry, we obtain a class of efficiencies showing universality upto second order in Carnot value. Interestingly, CA-efficiency emerges as the lower bound of this class (see Eq. (\ref{calow})).
 
Let us first consider the work extraction by reversible means. We denote the energy and entropy of the system by $(U_h, S_h)$ and $(U_c, S_c)$,  at temperature $T_h$ and $T_c$, respectively. Also, let $C_{V}^{} (T)$ be the heat capacity at constant volume of the system as a function of its temperature $T$. For simplicity, we take the volume of the system to be fixed during the cycle. In the first stage, the system, initially at temperature $T_c$, is coupled with the hot reservoir through a reversible  heat engine. The engine extracts work via infinitesimal heat cycles which successively increase the temperature of the system until it attains thermal equilibrium with the hot reservoir. The maximum extracted work, also called exergy \cite{mor}, in the first stage, is then $W_1 = T_h \Delta S - \Delta U$. Here, $\Delta X = X_h-X_c$. In the second stage, the system is coupled to the cold reservoir via a reversible heat engine. Again, work can be obtained till thermal equilibrium is achieved with the cold reservoir, completing one cycle for the auxiliary system. So, the exergy in the second stage is $W_2 = \Delta U - T_c \Delta S$.  The total work extracted in one cycle is $W_0 \equiv W_1 + W_2 = \Delta T \Delta S$. The input heat extracted from the hot reservoir is $Q_h = T_h \Delta S$. So, the efficiency $\eta = W_0/Q_h$,  as expected, is equal to the Carnot value. Note that, since the auxiliary system interpolates between two given initial and final equilibrium states, so $\Delta U$ and $\Delta S$ are given parameters of the cycle.

\textit{Linear irreversible heat engine}. Such an engine is in simultaneous contact with both the heat source and sink (see Fig. 1), and it can be analyzed within the linear irreversible framework \cite{ons226,cal}. In the first stage, heat flux $\dot{Q}_h$ enters the engine from the hot reservoir, power $P$ is generated, and a heat flux $\dot{Q}_c$ goes into the finite sink whose instantaneous temperature $T$ satisfies: $T_c \leq T \leq T_h$. 
\begin{figure}[bp]
	\includegraphics[scale=0.425]{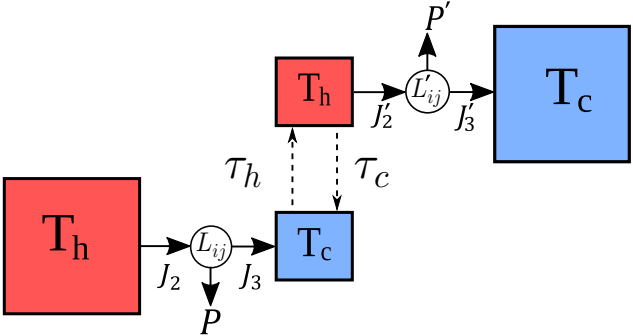}
	\caption{The schematic diagram of the two-stage heat engine. In the first stage, the state of the auxiliary system (smaller box), connected via a linear irreversible engine with an infinite reservoir, changes from one at temperature $T_c$ to $T_h$, in time $\tau_h$. In the second stage that lasts for time $\tau_c$, the system comes back to the initial state at temperature $T_c$ via a similar engine, thus completing a cycle.}
	\label{fig:fin}
\end{figure}
Let, at some instant, $dW$ be the infinitesimal work performed by the engine on the environment against a uniform external force $F$. Then,  $dW = -F\,dx$, where $x$ is the conjugate thermodynamic variable of $F$. The power output is $P = -F\,\dot{x} = \dot{Q}_h - \dot{Q}_c$. The instantaneous total entropy production rate in the reservoirs \cite{van190,izu180}:
\begin{eqnarray}\label{b}
\dot{S} &=& \frac{\dot{Q}_c}{T} -\frac{\dot{Q}_h}{T_h}, \\
 &=&\frac{F\,\dot{x}}{T}+\dot{Q}_h\left(\frac{1}{T}-\frac{1}{T_h}\right).
\end{eqnarray}   
Here, we can identify two thermodynamic force-flux pairs: $X_1\equiv F/T$, $J_1\equiv\dot{x}$, $X_2 \equiv 1/T-1/T_h$, $J_2\equiv\dot{Q}_h$  \cite{van190}.
Then, we can write: $\dot{S}= J_1X_1+J_2X_2$, and $P=-TJ_1X_1$. If the thermodynamic forces are small, then the fluxes can be expressed in the following form \cite{ons226,cal}:
\begin{equation}\label{c}
J_1=L_{11}X_1+L_{12}X_2,
\end{equation}
\begin{equation}\label{d}
J_2=L_{21}X_1+L_{22}X_2,
\end{equation}
where $L_{i j} (i,j=1,2)$ are Onsager coefficients satisfying the reciprocity relation $L_{12} = L_{21}$, and  $L_{12} L_{21} \leq  L_{11}L_{22}$. Using Eq. (\ref{c}), the power output can be expressed in terms of $J_1$ as 
\begin{equation}\label{e}
P=\frac{L_{12}}{L_{11}} T X_2 J_1-\frac{T}{L_{11}}J_1^{2}.
\end{equation}
The input heat flux becomes 
\begin{equation}\label{f}
J_2=\frac{L_{12}}{L_{11}} J_1+L_{22}(1-q^{2})X_2,
\end{equation}
where the coupling parameter, $q^{2}\equiv L_{12}^{2}/(L_{11}L_{22})$, satisfies $0\leq q^{2}\leq 1$ \cite{ked189}. Using Eqs. (\ref{e}) and (\ref{f}), we can write 
 $\dot{Q}_c\equiv J_3=J_2-P$ as 
\begin{equation}\label{g}
J_3=\frac{L_{12}T}{L_{11}T_h} J_1+L_{22}(1-q^{2})X_2+\frac{T}{L_{11}} J_1^{2}.
\end{equation}
Now, the rate of change of the sink temperature is given by $C_{V}^{}\dot{T}=J_3$, where $\dot{T} >0$. Solving for $J_1$ from Eq. (\ref{g}) and substituting in Eq. (\ref{f}), we get the input heat flux as a function of the form $J_2 (T, \dot{T})$. Here, we discuss only the tight-coupling condition, $q^{2}=1$, which is analytically amenable. Finally, we obtain    
\begin{equation}\label{h}
J_2=\frac{L_{22}}{2T_h}\left(\sqrt{1+\frac{4T_h^{2}C_{V}^{}\dot{T}}{L_{22}T}} -1\,\right).
\end{equation}
We assume that the finite-size sink takes time $\tau_h$ to reach thermal equilibrium with the hot reservoir. The work output in the first stage is given by
\begin{equation}\label{i}
W=\int_{0}^{\tau_h}J_2 dt-\int_{0}^{\tau_h}J_3dt=\int_{0}^{\tau_h}J_2\, dt- \Delta U,
\end{equation}
where $\Delta U = {\displaystyle \int_{T_c}^{T_h} } { C_{V}^{}} \dot{T} \; dt$.

\textit{Optimization.} In order to maximize $W$, the input heat  $Q_h \equiv \int_{0}^{\tau_h}J_2 dt$ needs to be maximized. The corresponding Euler-Lagrange (EL) equation for the flux $J_2$ is given by
\begin{equation}\label{j}
\frac{d}{dt}\left(\frac{\partial J_2}{\partial\dot{T}}\right)-\frac{\partial J_2}{\partial T}=0.
\end{equation}
Now, the Onsager coefficient $L_{22}$ and heat capacity $C_{V}^{}$ may depend on the temperature. Even the above EL equation seems difficult to solve. To simplify, we assume $\frac{4T_h^{2}C_{V}^{}\dot{T}}{L_{22}T}\ll1$ (see A of supplemental material for justification of this approximation). Then, the input heat flux is approximated up to second order as 
\begin{equation}\label{k}
J_2\approx C_{V}^{}T_h\frac{\dot{T}}{T}-\frac{C_{V}^{2} T_h^{3}}{L_{22}}\left(\frac{\dot{T}}{T}\right)^{2}.
\end{equation}
The EL equation yields 
\begin{equation}\label{l}
\frac{2C_{V}^{2}}{L_{22}}\frac{\ddot{T}}{T^{2}} +
\frac{\partial}{\partial T}\left(\frac{C_{V}^{2}}{L_{22}T^{2}}\right)\dot{T}^{2}=0.
\end{equation}
Multiplying Eq. (\ref{l}) with $\dot{T}$ and simplifying, we can get
\begin{equation}
\frac{d}{dt}\left(\frac{C_{V}^{2} \dot{T}^{2}}{L_{22} T^{2}}\right)=0.
\end{equation}
Integrating the above equation w.r.t time, we get
\begin{equation}\label{m}
\frac{C_{V}^{} \dot{T}}{\sqrt{L_{22}} T}= A,
\end{equation}
where $A>0$ is a constant of integration. Then, integrating Eq. (\ref{m}) with respect to time, we get $A=B/\tau_h$, where $B = {\displaystyle \int_{T_c}^{T_h} }\frac{C_{V}^{}}{\sqrt{L_{22}}\,T}dT >0 $. The input heat becomes
\begin{eqnarray}\label{n}
Q_h &=& T_h \int_{T_c}^{T_h}\frac{C_{V}^{} dT}{T}-T_h^{3} A^{2} \int_{0}^{\tau_h} dt, \label{n} \\
&=& T_h\Delta S- \frac{T_h^{3} B^{2}}{\tau_h}, \label{o}
\end{eqnarray}
where ${\displaystyle \int_{T_c}^{T_h} } \frac{C_{V}^{}}{T} \; dT = \Delta S >0$ is the entropy injected into the system in the first stage. Then, Eq. (\ref{i})  is evaluated to be
\begin{eqnarray}\label{p}
W &=& T_h\Delta S  - \frac{T_h^{3} B^{2}}{\tau_h} -\Delta U, \\
  &\equiv& W_1 -  \frac{T_h^{3} B^{2}}{\tau_h}.
\end{eqnarray}
Clearly, as the time duration of the first stage diverges, the work output approaches its maximal value $W_1$. Upon rewriting Eq. (\ref{o}) as
\begin{equation}\label{o2}
\Delta S - \frac{Q_h}{T_h}  = \frac{T_h^{2} B^{2}}{\tau_h} >0,
\end{equation}
we note that the left-hand side of the above equation is the sum of the entropy changes in 
the sink and the hot reservoir, and so, it is equal to the total entropy generated
due to the engine. In fact, the inverse proportionality of this quantity on the duration 
of the process, is the basic assumption of the low-dissipation model \cite{esp150}.

In the second stage, the heat engine is connected between the system at initial temperature $T_h$ (finite heat source) and the cold reservoir at temperature $T_c$. We assume that the source takes a given time $\tau_c$ to arrive at temperature $T_c$ (see Fig. 1). The rate of change of the source temperature can be written as $-C_{V}^{}\dot{T}^{'}=J_2^{'}$, where $\dot{T}^{'} <0$ denotes the decreasing temperature of the source. Thus, the heat absorbed from the source  in the second stage is $Q_h^{'} = \Delta U$. The optimized heat rejected to the cold reservoir (that maximizes the work extracted in the second stage) can be calculated along similar lines (see B of supplemental material):
\begin{equation}\label{v}
Q_c^{'}=T_c \Delta S+ \frac{T_c  B^{'2}}{\tau_c},
\end{equation}
where $B^{'} = {\displaystyle \int_{T_c}}^{T_h} \frac{C_{V}^{}}{\sqrt{L_{22}^{'}}} dT^{'}$.
Thus, the maximum work output in the second stage  is given by \cite{izu180}:
\begin{eqnarray}\label{w}
W^{'} &=& \Delta U  - T_c \Delta S- \frac{T_c  B^{'2}}{\tau_c}, \\
&\equiv& W_2 -  \frac{T_c  B^{'2}}{\tau_c}.
\end{eqnarray}
From Eq. (\ref{v}), we note that $Q_c^{'}/T_c$ is the entropy added reversibly to the cold reservoir, while $\Delta S$ is the change (decrease) in the entropy of the finite source. Thus, $Q_c^{'}/T_c - \Delta S = B^{'2}/\tau_c$, is identified as the entropy generated in the second stage, which is also inversely proportional to the time $\tau_c$ spent on the process. Thus, we conclude that our model in which the individual stages are optimized for maximum work extraction with given times, yields the low-dissipation behavior \cite{esp150}. This is our first main result.

The total extracted work, $W_{\rm tot}=W+W^{'}$, is given by 
\begin{equation}\label{x}
W_{\rm tot}= \Delta T \Delta S-\frac{T_h^{3}B^{2}}{\tau_h}- \frac{T_c B^{'2}}{\tau_c}.
\end{equation} 
Now, the cycle lasts for a time $\tau=\tau_h+\tau_c$. Therefore, the power generated per cycle is  $P=W_{\rm tot}/\tau$. Maximizing the power  with respect to $\tau_h$ and $\tau_c$ 
(for fixed values of $T_h$, $T_c$, and $\Delta S$ \cite{esp150}), we obtain the optimal allocation of times as:
\begin{equation}
\tau_h^{\star}=\frac{2T_h^{3}B^{2}}{\Delta T \Delta S}
\left(1+ \Gamma \sqrt{\frac{T_c}{T_h}}\right),
\end{equation}
\begin{equation}
\tau_c^{\star}=\frac{2 T_c B^{'2}}{\Delta T \Delta S}
\left(1+\frac{1}{\Gamma} \sqrt{\frac{T_h}{T_c}}\right),
\end{equation}
where $\Gamma = B^{'}/(T_h B)$. 
The efficiency at maximum power:
\begin{equation}
\eta_{\rm MP}={\eta_{\rm C}} \left({2-\frac{\eta_{\rm C}}
	{1+ \Gamma\sqrt{1-\eta_{\rm C}}}} \right)^{-1}.
\label{emax}
\end{equation}
We note that $\Gamma$ is determined by two kinds of control parameters: the thermostatic property (heat capacity) of the auxiliary system and thermal conductivities of the contacts with reservoirs (through $L_{22}$ and $L_{22}^{'}$). To analyze the dependence  on 
these factors, we consider the following case.

\begin{figure}[h]
	\includegraphics[scale=0.56]{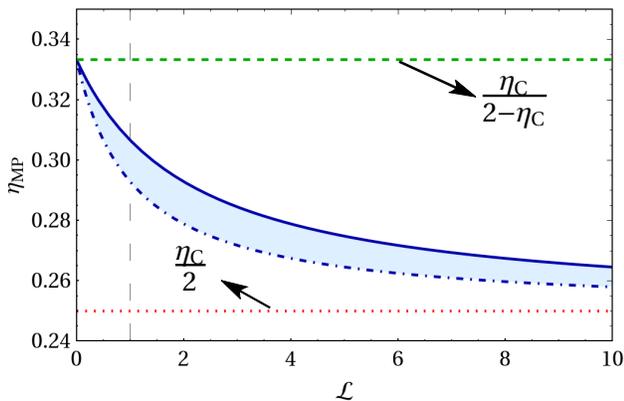}
	\caption{\label{fig:lplot} The bounds of $\eta_{\rm MP}$ as given in Eq. (\ref{Lb}) (shaded region) are  plotted as a function of $\mathcal{L}$, for $\eta_{\rm C}=0.5$. The upper (dashed) and the lower (dotted) horizontal lines are the bounds given in Eq. (\ref{ul}). The vertical large-dashed line represents the symmetric case $\mathcal{L}=1$.}
\end{figure}

\textit{Special case.} In the near-equilibrium linear regime, we may take $L_{22}$ and $L_{22}^{'}$ to be constant parameters  (independent of temperature) \cite{kon}. Then, we denote $\mathcal{L} \equiv \sqrt{L_{22}/L_{22}^{'}}$, and $T_m = {\Delta U}/{\Delta S}$, so that $\Gamma = \mathcal{L} T_m/T_h$. From the mathematical properties of the function $U(S)$ and the mean-value theorem \cite{joh012}, we know that $T_m$ is bounded as: $T_c < T_m < T_h$. For a given $\mathcal{L}$, as $T_m \to T_c$ ($T_m \to T_h$), we obtain the upper (lower) bounds for $\eta_{\rm MP}$: 
\begin{equation}\label{Lb}
\frac{\eta_{\rm C}}{2-
	\frac{\eta_{\rm C}}{1+ \mathcal{L}\sqrt{1-\eta_{\rm C}}}} <  \eta_{\rm MP} < \frac{\eta_{\rm C}} {2-
	\frac{\eta_{\rm C}}{1+ \mathcal{L}\, (1-\eta_{\rm C})\sqrt{1-\eta_{\rm C} }}}.
\end{equation}
On the other hand, the parameter $\mathcal{L}$ can take values in the range $0 < \mathcal{L} < \infty$. Further, $\eta_{\rm MP}$ decreases monotonically as $\mathcal{L}$ increases. 
Due to these conditions, the global bounds on EMP are as follows:
\begin{equation}\label{ul}
\frac{\eta_{\rm C}}{2} < \eta_{\rm MP} < \frac{\eta_{\rm C}}{2-\eta_{\rm C}}, 
\end{equation}
where the lower (upper) bound is approached when $L_{22} \gg L_{22}^{'}$ or $\mathcal{L} \to \infty$ ($L_{22} \ll L_{22}^{'}$ or $\mathcal{L} \to 0$). The bounds from Eqs. (\ref{Lb}) and (\ref{ul}) are plotted  in Fig. \ref{fig:lplot}.

\begin{figure}[hpt!]
\includegraphics[scale=0.532]{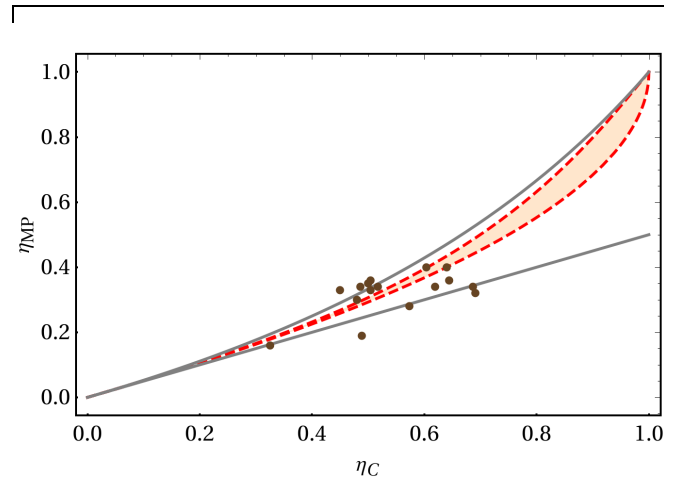}
\caption{\label{fig:eta}Bounds on EMP of the two-stage heat cycle plotted against Carnot value $\eta_{\rm C}$. Upper and lower solid curves are as given in Eq. (\ref{ul}). The dashed lines depict the bounds of $\eta_{\rm MP}^{({\rm sym})}$ as in Eq. (\ref{calow}), the lower being the CA-efficiency.
The shaded region defines efficiencies under the symmetry condition $L_{22} = L_{22}^{'}$, which show universality up to second order in  $\eta_{\rm C}$. The dots represent data on the observed efficiencies of a few thermal and nuclear plants as in Ref. \cite{joh489}.}
\end{figure}

In particular, we consider the symmetric situation, $L_{22}=L_{22}^{'}$, or $\mathcal{L} =1$.
In this case, as $T_m \to T_c$ ($T_m \to T_h$), we obtain the upper (lower) bound of the EMP (see C of supplemental material for the bounds on $\Gamma$), given by:
\begin{equation}\label{calow}
 1-\sqrt{1-\eta_{\rm C}} <  \eta_{\rm MP}^{({\rm sym})} < {\eta_{\rm C}} \left({2-
 \frac{\eta_{\rm C}}{1+  (1-\eta_{\rm C}) \sqrt{1-\eta_{\rm C}}}}\right)^{-1},
\end{equation}
which is our second main result. The above bounds are depicted in Fig. \ref{fig:eta} (shaded region). We note that, in the symmetric case, we have a class of efficiencies which satisfy the criterion of universality upto second order, i.e. $\eta_{\rm MP}^{({\rm sym})} = {\eta_{C}}/{2} + {\eta_{C}^{2}}/{8}+ O[{\eta_{C}^{3}}]$. Interestingly, the CA-efficiency emerges as the lower bound for EMP in the symmetric case. Finally, to find the impact of the auxiliary system on $\eta_{\rm MP}$, we assume the following general form of $C_V=\beta\, T^{\alpha}$, where $\alpha, \beta$ are constants. We get 
\begin{equation}
\eta_{\rm MP}=\frac{\eta_{\rm C}}{2-\eta_{\rm C}\left[1+\mathcal{L}\left(\frac{\alpha}{ \alpha+1}\frac{1-(1-\eta_{\rm C})^{\alpha+1}}{1-(1-\eta_{\rm C})^{\alpha}}\right)\sqrt{1-\eta_{\rm C}}\right]^{-1}
	}.
\end{equation}
Expanding the above $\eta_{\rm MP}$ in terms of $\eta_{\rm C}$, we obtain 
\begin{widetext}
	\begin{equation}
	\eta_{\rm MP} = \frac{\eta_{\rm C}}{2}+\frac{\eta_{\rm C}^{2}}{4(1+\mathcal{L})}+\frac{1+2\mathcal{L}}{8(1+\mathcal{L})^{2}} \eta_{\rm C}^{3} +\frac{6+
	(23- 2\alpha)\mathcal{L}(1+\mathcal{L})}
	{96(1+\mathcal{L})^{3}} \eta_{\rm C}^{4}+ \mathcal{O}[\eta_{\rm C}^{5}]. 
	\end{equation}
\end{widetext}
Thus, we notice that the efficiency at maximum power depends only weakly on the nature of the auxiliary system.

\textit{Conclusions.}
We have proposed a two-stage heat cycle with a finite time period, where an auxiliary system plays the role of a finite sink in one stage and a finite heat source in the other stage. In each stage, a linear irreversible engine, coupled to the finite system and one of the heat reservoirs, is optimized for maximum work output in a given time. The usual formulations of finite-time Carnot engines involve four stages where, for simplification, the time spent on adiabatic stages is assumed to be negligible \cite{che374,sch200,esp150}. In this regard,
the proposed simpler model does not involve any adiabatic stages. Also, the model does not depend on special conditions such as endoreversibility \cite{cur22,rub127}.

It is observed that under the tight-coupling condition, the present model is formally analogous to the low-dissipation model \cite{esp150}. In fact, the dissipation constants that are introduced phenomenologically in the latter model are determined within the theory in terms of heat transfer coefficients and heat capacity of the auxiliary system. It is to be noted that LD behaviour has been derived both within classical or quantum domains with mesoscopic or microscopic systems \cite{sch200,esp041,Cavina2017}. However, a basic derivation of this behavior for classical, macroscopic engines is not known. The  present model provides a foundation for the low-dissipation assumption for {\it macroscopic} thermal machines \cite{esp150}, based on notions from  linear irreversible thermodynamics.

Further, the efficiency at maximum power is found to be consistent with the universal properties of efficiency observed in other models. 
In contrast to the low-dissipation model where CA-efficiency is obtained
under the symmetric condition \cite{esp150}, here we get a class of efficiencies
with CA-efficiency being the lower bound. Curiously, we note that 
though the observed efficiencies of quite a few well-known plants fall within the bounds (Eq. (\ref{ul})), the available data seems to be excluded from the region of symmetry (shaded area in Fig. \ref{fig:eta}). Finally,  among other generalizations, it will be interesting to extend the model beyond the linear regime \cite{izu100,izu052}, as well as for refrigerators \cite{jim057}.

I. I. gratefully acknowledges IISER Mohali for the financial support, and  also thanks
Varinder Singh and Jasleen Kaur for valuable discussions and suggestions. R. S. J. would like to acknowledge Stefano Ruffo and Edgar Roldan for interesting discussions, and Rozario Fazio for hospitality at the Abdus Salam International Center for Theoretical Physics, Trieste, Italy, where a  part of the above work was done.

\end{document}


\title{Supplemental Material: Efficiency of a two-stage heat engine at optimal power}
\author{I. Iyyappan}
\email[]{iyyappan@iisermohali.ac.in}
\author{Ramandeep. S. Johal}
\email[]{rsjohal@iisermohali.ac.in}
\affiliation{Department of Physical Sciences,\\ Indian Institute of Science Education and Research Mohali,\\
Sector 81, S. A. S. Nagar, Manauli PO 140306, Punjab, India.}

\maketitle
\section*{A. Input heat flux approximation}
The input heat flux in the first stage is given by (Eq. (9))
\begin{equation}\label{m}
J_2=\frac{L_{22}}{2T_h}\left[\sqrt{1+\frac{4T_h^{2} C_V\dot{T}}{L_{22}T}}-1\right]. \tag{S1}
\end{equation}
Here, the Onsager coefficient $L_{22}\geq0$ and the rate of change of temperature of the finite-size 
sink, $\dot{T}\geq0$. When the quantity
$\frac{4T_h^{2} C_V\dot{T}}{L_{22}T}\ll1$, the input heat flux can be approximated as 
\begin{equation}\label{k}
J_2\approx C_{V}^{}T_h\frac{\dot{T}}{T}-\frac{C_{V}^{2} T_h^{3}}{L_{22}}
\left(\frac{\dot{T}}{T}\right)^{2},\tag{S2}
\end{equation}
which gives the following optimal rate of change for the temperature (Eq. (15)) 
\begin{equation}
\frac{C_V\dot{T}}{\sqrt{L_{22}}T}=A=\frac{B}{\tau_h}. \tag{S3}
\end{equation}
Therefore, the quantity
\begin{equation}
\frac{4T_h^{2} C_V\dot{T}}{L_{22}T}=\frac{4T_h^{2}B}{\sqrt{L_{22}}\tau_h}. \tag{S4}
\end{equation}
For a constant $L_{22}$, we get $B=\Delta S/\sqrt{L_{22}}$, where $\Delta S\equiv\int_{T_c}^{T_h}(C_{V}^{}/T)\; dT$. Rewriting the above equation
\begin{equation}\label{s5}
\frac{4T_h^{2} C_V\dot{T}}{L_{22}T}=\frac{4T_h^{2}\Delta S}{L_{22}\tau_h}. \tag{S5}
\end{equation}
Thus, our approximation implies the following condition:
\begin{equation}
\frac{4T_h^{2}\Delta S}{L_{22}\tau_h} \ll 1, \tag{S6}
\end{equation}
which requires a sufficiently long duration for the first stage, given 
the values of other parameters.

Further, the consistency of our approximation can be checked with 
the optimal power solution, for which we obtained
the optimal time $\tau_h$ as
\begin{equation}
\tau_h^{\star}=\frac{2T_h^{3}B^{2}}{\Delta T\Delta S}(1+\Gamma\sqrt{\theta})=\frac{2T_h^{3}\Delta S}{L_{22}\Delta T}(1+\Gamma\sqrt{\theta}), \tag{S7}
\end{equation}
where $\Gamma\equiv B^{'}/(T_h B)$ and $\theta\equiv T_c/T_h$. 
Substituting the above $\tau_h^{\star}$ in Eq. (\ref{s5}), we get
\begin{equation}
\frac{4T_h^{2} C_V\dot{T}}{L_{22}T}=\frac{2\Delta T}{T_h}\frac{1}{(1+\Gamma\sqrt{\theta})}. \tag{S8}
\end{equation}
Since $\Delta T\ll T_h$ in the linear-irreversible regime, it follows that
\begin{equation}
\frac{4T_h^{2} C_V\dot{T}}{L_{22}T}\ll 1. \tag{S9}
\end{equation} 
\section*{B. Optimization in the second stage}
Here, we briefly review the work of Izumida and Okuda for completeness \cite{izu180}. The total entropy production rate at the finite-size hot source and the cold reservoir is given by \cite{izu180}
\begin{equation}\label{b}
\dot{S}^{'}=-\frac{\dot{Q}_h^{'}}{T^{'}}+\frac{\dot{Q}_c^{'}}{T_c}=\frac{F^{'}\,\dot{x}^{'}}{T_c}+\dot{Q}_h^{'}\left(\frac{1}{T_c}-\frac{1}{T^{'}}\right). \tag{S10}
\end{equation}   
Defining the pairs of thermodynamic forces and fluxes as $X_1^{'}\equiv F^{'}/T_c$, $J_1^{'}\equiv\dot{x}^{'}$, $X_2^{'} \equiv 1/T_c-1/T^{'}$, $J_2^{'}\equiv\dot{Q}_h^{'}$. In the linear response regime, the fluxes can be written as
\begin{equation}\label{c}
J_1^{'}=L_{11}^{'}X_1^{'}+L_{12}^{'}X_2^{'},  \tag{S11}
\end{equation}
\begin{equation}\label{d}
J_2^{'}=L_{21}^{'}X_1^{'}+L_{22}^{'}X_2^{'}, \tag{S12}
\end{equation}
where $L_{i j}^{'}(i,j=1,2)$ are Onsager coefficients with $L_{12}^{'}=L_{21}^{'}$. The power output becomes, $P^{'}=-T_cJ_1^{'}X_1^{'}$. Using the Eq. (\ref{c}), the power output can be expressed in terms of $J_1^{'}$ as 
\begin{equation}\label{e}
P'=\frac{L_{12}^{'}}{L_{11}^{'}}\left(1-\frac{T_c}{T^{'}}\right) J_1^{'}-\frac{T_c}{L_{11}^{'}}J_1^{'2}. \tag{S13}
\end{equation}
The input heat flux becomes 
\begin{equation}\label{sj2}
J_2^{'}=\frac{L_{12}^{'}}{L_{11}^{'}} J_1^{'}+L_{22}^{'}(1-q^{'2})X_2^{'}, \tag{S14}
\end{equation}
where $q^{'2}\equiv L_{12}^{'2}/(L_{11}^{'}L_{22}^{'})$ with $0\leq q^{'2}\leq1$. The ejected heat flux to the cold reservoir, $\dot{Q}_c^{'}\equiv J_3^{'}=J_2^{'}-P^{'}$. Using Eqs. (\ref{e}) and (\ref{sj2}), we get the ejected heat flux as 
\begin{equation}\label{sj3}
J_3^{'}=\frac{L_{12}^{'}T_c}{L_{11}^{'}T^{'}} J_1^{'}+L_{22}^{'}(1-q^{'2})X_2^{'}+\frac{T_c}{L_{11}^{'}} J_1^{'2}. \tag{S15}
\end{equation}
The rate of change of the source temperature is given by $J_2^{'}=-C_V\dot{T}^{'}$, where 
 $\dot{T}^{'} < 0$ indicates the decreasing source temperature. Substituting the above $J_2^{'}$ in Eq. (\ref{sj2}), we get  $J_1^{'}$ under the tight-coupling condition ($q^{'2}=1$) as 
\begin{equation}\label{sj1}
J_1'=-\frac{L_{11}^{'}}{L_{21}^{'}}C_V\dot{T}^{'}. \tag{S16}
\end{equation}
Substituting the above $J_1^{'}$ in Eq. (\ref{sj3}), we obtain the output heat flux with the tight-coupling condition as 
\begin{equation}
J_3^{'}=-T_cC_V\frac{\dot{T}^{'}}{T^{'}}+\frac{T_cC_V^{2}\dot{T}^{'2}}{L_{22}^{'}}. \tag{S17}
\end{equation}
Optimizing the $J_3^{'}$ by using the EL equation, we get 
\begin{equation}\label{el}
\frac{2C_V^{2}\ddot{T}^{'}}{L_{22}^{'}}+\dot{T}^{'2}\frac{\partial}{\partial T^{'}}\left(\frac{C_V^{2}}{L_{22}^{'}} \right)=0. \tag{S18}
\end{equation}
Multiplying Eq. (\ref{el}) with $\dot{T}^{'}$ and simplifying further, we get 
\begin{equation}
\frac{d}{dt}\left(\frac{C_V^{2}\dot{T}^{'2}}{L_{22}^{'}} \right)=0\tag{S19}
\end{equation} 
Integrating the above equation, we get 
\begin{equation}\label{sac}
\frac{C_{V}\dot{T}^{'}}{\sqrt{L_{22}^{'}}}=- A^{'}, \tag{S20}
\end{equation}
where $A^{'}>0$ is the integration constant. Integrating Eq. (\ref{sac}) with respect to time, we get $A^{'}=B^{'}/\tau_c$, 
where $B^{'}\equiv \int_{T_c}^{T_h}\frac{C_{V}}{\sqrt{L_{22}^{'}}}dT^{'}$.
Now, the optimized ejected heat becomes
\begin{equation}
Q_c^{'}=-T_c \int_{0}^{\tau_c}\frac{C_{V}\dot{T}^{'} }{T^{'}}dt+T_c A^{'2} \int_{0}^{\tau_c} dt
 = T_c\Delta S+\frac{T_cB^{'2}}{\tau_c}. \tag{S21}
\end{equation}
\section*{C. Bounds on parameter $\Gamma$ for the symmetric case}
Let us consider the inequality 
\begin{equation}
m\int_{a}^{b}g(x)dx\leq\int_{a}^{b}f(x)g(x)dx\leq M\int_{a}^{b}g(x)dx, \tag{S23}
\end{equation}
with $m\leq f(x)\leq M$. Assuming the functions $f(x)$ and $g(x)$ to be integrable between the limits $a$ and $b$, we take the symmetrical situation where $L_{22}(T)=L_{22}^{'}(T^{'})$ and $T_c\leq T\leq T_h$. Therefore, we can write
\begin{equation}
\frac{1}{T_h}\int_{T_c}^{T_h}\frac{C_V dT}{\sqrt{L_{22}}}\leq\int_{T_c}^{T_h}\frac{C_V dT}{\sqrt{L_{22}}T}\leq \frac{1}{T_c}\int_{T_c}^{T_h}\frac{C_V dT}{\sqrt{L_{22}}}. \tag{S24}
\end{equation}
Using the definitions of $B$ and $B^{'}$, we get
\begin{equation}
\frac{B^{'}}{T_h}\leq B \leq \frac{B^{'}}{T_c}. \tag{S25}
\end{equation}
Rewriting the above equation, we get the bounds as
\begin{equation}
\frac{T_c}{T_h} \leq \Gamma \leq 1.
\tag{S26}
\end{equation}